# Reliability Testing Strategy

Reliability in Software Engineering

Kevin Taylor-Sakyi
Aston University
Engineering & Applied Science
Birmingham, UK
Kevin.sakyi@gmail.com
www.kevintaylorsakyi.me

**Abstract— *This paper presents the core principles of reliability in software engineering; outlining why reliability testing is critical and specifying the process of measuring reliability. The paper provides insight for both novice and experts in the software engineering field for assessing failure intensity as well as predicting failure of software systems. Measurements are conducted by utilizing information from an operational profile to further enhance a test plan and test cases – all of which this paper demonstrates how to implement.***

## I. Introduction

Amazon, aerospace, and healthcare systems – what is the underlying factor of these systems? The importance of their reliability! Prominence of software systems in this *Information Age* is becoming more vital for operations within organizations; though these systems demonstrate competitive advantages and the like, *every rose has its thorns*. According to [1] it was estimated that less than 5% of testers were competent in in utilizing models to predict software reliability in the late 1990s; though this measurement is outdated it's a good indication that a gray area in dealing with reliability in testing strategies is still evident. This report delivers reliability theory in relation to software engineering, how it's measured & calculated, and how to develop a test plan to assess the reliability of a software system.

## II. Software Reliability

### A. What is it?

Software reliability is "the probability of failure-free operation of a computer program for a specific time in a specific environment" [2]. In other words, creating a test that produces identical reliability measures results repeatedly; *failure* meaning "the program in its functioning has not met user requirements in some way" [3]. For example, a pacemaker monitoring software that alerts a doctor of a patients heart condition is considered to have failed if accurate information is not delivered to the doctor within a specified time constraint; which in turn expresses a relatively low reliability of the system. In 2014, the Heart Rhythm Society mentioned that approximately 600,000 individuals *are implanted with pacemakers each year* globally [4]. If the reliability rate of these systems is deemed to be 45% for 600,000 individuals, the chance of survival is a mere 45%, contingent upon doctors not notified within a certain timeframe.

Reliability testing efforts can be said to predict failures that are likely to happen in specified system operations, identifying areas of which faults that need the most efforts to fix. These are typically categorized into *measurements*, *models,* and *methods of quality improvements.*

### B. Why measure reliability?

Reliability has always been centered on computer hardware, *i.e. how durable is a component of a printer, keyboard, etc. yet,* the same cannot be said about software systems. Thus standards of measuring reliability of hardware cannot be utilized as they focus on "wearing out processes" [6], unlike software, which does not erode or wear out.

Testing of software systems has had its fair share of interest in industry, however there's a gray area of concern. The complexity of software systems increase so does the acceptable definition of reliability, presently not commercially agreed upon with regards to measurement techniques. Testing of software, currently practiced in many developmental environments merely validates if a system or product meets business requirements but not how reliable the product is.

Software's that are considered reliable based on the *software requirements engineering* (SRE) process cannot only save lives, but can also bring profits to organization such as Amazon; or increase the credibility of a critical system such as an aerospace's traffic monitoring system. Proving the necessity of being a phase within developmental processes – separate from standard software testing. This supplementary ad-hoc process does not replace current processes for testing software, but allows *precise decisions-making* during software development and allows "everyone more concretely aware of SR by focusing attention on it" [3]; promoting means of reducing software developmental and maintenance costs.

*C. Engineering Process*

In ensuring reliability of a system a systematic approach must be followed to ensure a *safe, correct,* and functional software that meets the *operational aspects of usability and user-friendliness* [3]. The engineering process is impartial to developmental methods (*i.e. waterfall, agile, etc.*) however the process may invoke changing designs, frameworks and the like to produce a system with greater reliability. The engineering process is as follows:

*1) **Defining the product**:* involves depicting actors (users, suppliers, customers, etc.) of a system and determining the base product and its accompanying systems and different variations to establish which tests are suitable for each component

   *a)* Refering to *pacemaker system x* ('*x*' denoting a generic pacemaker system that informs doctors of patients heart rates, etc.) mentioned earlier, the different variations entailing measurements of heart rhythm, transfering of measurements, etc. – recognizing these variations allow different types of tests to be implemented. Promoting test types to be specialized per variation.

*2) **Implementing operational profiles**:* Identifying major tasks to be accomplished by the system (customers, users, etc.) and their rate of occurrence, they must preserve control until task has been completed [3]. These profiles facilitate testing to be conducted efficiently, allowing tracability of "the reliability being achieved" [5]. Development of an operational profile consists of the following:

   *a)* Identify initiators of operations: Indicating different users or user types of the system, typically found through analyzing the pre-described customer types during the inception phase of the requirements analysis (For example, in the case of pacemaker system *x* a doctor viewing a weekly generated report via email would not be considered an initiator of that variation; manually searching for inconsistencies within a set-time frame however makes the doctor an initiator of this variation). Figure 2 expresses possible initiators within this pacemaker system.

   *b)* Creation of operations list: Operations are jobs conducted within the software system – these are derived from system requirements (functional & non-funcational), diagrams (*i.e. activity diagram*), and discussions with the various user types. Involving expected users occasionally highlights areas neglected during requirements gathering. Refer to Figure 3.

   *c)* Review operations list: Consists of amending list to ensure high probability, should consider view points of experts within initiators. Resulting in merging opeartions to allow a *system test* or partitioning of operations to permit selective testing resources.

   *d)* Determining occurence rates: "The number of operations divided by the time the total set of operations is running" [3]. Potentially obtained by examining existing data, business data obtained from marketers, estimating using the Delphi method with various experts involved, and lastly, due to its cost, manually calculating estimates. Refer to Figure 4.

   *e)* Determining occrence probabilities: Dividing each operation's occurrence rate by the sum of the operation occurrence rate. Refer to Figure 4.

*3) **Engineering the reliability**:* Specifying the *just right* goals of meeting the reliability objectives, demands defining the folling within specific system [3]:

   *a) Define meaing of failure*

   *b) Indicate common unit for all failure intensities-allows*

   *c) Establish failure intensity objective for each associated system (operations, variation, etc.)*

   *d) Locate failure intensity objective for whole system & select strategies (models, etc.) that "optimally meet the developed software failure intensity objective"*

*4) **Preparing for test**:* Incorporates test cases and procedures which pilot "feature, load, and regression tests" [3]. Feature consists of independent tests on operations to determine if operations perform accurately. Load iterates large amounts of tests with confidence to imitate failure which may occur due to interactions between different operations. Regression test is done periodicly [3] and involves repetitive feature test after each build to determine failure based on amendments on the software system.

*5) Executing test:* Identification of failures, when they occur, and how severe they impact the system is found in this step [5]. You may use SRE to estimate & track failure intensity in this process to help remove failures.

*6) Guiding test:* Gathers all data relating to failed tests occured in testing to assist in the following decisions: tracing reliabilty growth, preparing acceptence testing, acceptance/rejection of a "supersystem", and releasing of a product entialing of all variations [3].

*7) Post Delivery & Maintenace life-cycle phase:* Phase which realibility is attained and the operational profile is experienced.

*D. Measuring & Calculating Reliability*

Establishing the SRE process is a good foundation, however the nature of quantitatively measuring software systems requires greater insight. Firstly, classification of the various types of failures must be determined, as reliability is concerned with the occurrence of different failures. This classification provides an orderly means of counting failures. Microsoft states that failures should be segmented into three groups, *unplanned events, planned events,* and *configuration failures* [12]. Figure 5 displays a breakdown of these categories appropriately.

Secondly, obtain failure data. As mentioned in the SRE process there are various means of obtaining the rates of failures – however failures must be documented from a users perspective; *i.e. allowing a user to report a failure*. The concern with this approach is that users often avoid reporting or fix faults themselves [12]. An alternative approach recommended by Jalote is by using polling methods in periodically asking users of a system to report any errors within specific operations.

Standards to be accounted for when measuring & calculating reliability should identify the following [8]:

- Fault introduction – defect in a software caused by altering or inserting new code
- Fault removal – *debugging* actions to remove identified faults
- Execution time ($\tau$) – duration of system running
- Mean failures experienced ($\mu$) – failures in specified time period
- Mean Time To Failure (MTTF) – measures the length in time that a system lasts in operation; MTTR $\approx 1/\lambda$
- Mean Time To Repair (MTTR) – measures average time taken to repair a failure
- Mean Time Between Failure (MTBF) – measurement of how reliable a component of a system is; MTBF $\approx \tau / \lambda$ or "the sum of mean time to failure (MTTF) and mean time to repair (MTTR)" [3]
- Converting $\lambda$ to reliability (R) – $R \approx \exp(-\lambda\tau)$ if $\lambda\tau$ is less than 0.05 then $R \approx 1 - \lambda\tau$

Identifying *failure intensity ($\lambda$)* – occurrence of failure within a specified time unit is recognized as a significant method in expressing software reliability [2]. Figure 6 shows a sample reliability model with the above standards incorporated within it. Viewing the model, it's evident that *failure rate* decreases as time of testing passes – displaying reliability growth. The model also illustrates that reliability of software systems "stays constant over time if no changes are made to the code or to the environmental conditions including the user behavior" – unlike hardware [10].

Failure intensity is typically visualized using models that effectively illustrate the failures experienced over time; this report demonstrates the use of the Basic Time Execution (BET) model and briefly mentions details of the Logarithmic-Poisson Execution Time (LPET) model. Both operate on the assumption that reliability testing utilizes operational profiles "and that every detected failure is immediately and perfectly repaired" [11].

Within the BET model (prediction model), failure measurement is determined using execution time (CPU or processor time). The following can be calculated accordingly [13]:

$$\mu(\tau) = \nu_0 \times \left(1 - e^{-\frac{\lambda_0}{\nu_0}\tau}\right)$$

Figure 1 – Mean failures experienced per time $\tau$

The following must be noted in Figure 1:

- $\lambda_0$ – initial failure intensity at beginning of execution
- $\nu_0$ – total number of failures over an endless time period from the beginning of system testing

$$\lambda(\tau) = \lambda_0 \times e^{-\frac{\lambda_0}{\nu_0}\tau}$$

Figure 1.1 – Failure Intensity for a given execution time

Computing the projected number of failures (Δμ) as well as the projected execution time (Δτ) to reach the failure intensity objective [11] may then be calculated using the formula shown in Figures 1.3 & 1.4. Where $\lambda_1$ is the current failure intensity and $\lambda_2$ is the failure intensity objective [13]. This model in effect allows the current and future reliability of software to be derived and assessed as a function a specified time-unit.

$$\lambda(\mu) = \lambda_0 \times \left(1 - \frac{\mu}{\nu_0}\right)$$

Figure 1.2 – Failure intensity λ in terms of μ

$$\Delta\mu = \frac{\nu_0}{\lambda_0} \times (\lambda_1 - \lambda_2)$$

Figure 1.3 – Expected number of failures & Execution time to reach the objective

$$\Delta\tau = \frac{\nu_0}{\lambda_0} \times \ln\left(\frac{\lambda_1}{\lambda_2}\right)$$

Figure 1.4 – Additional Execution time to reach the objective

Similar to the BET model, the LPET model uses execution time as a time unit yet it also includes calendar time (*not requiring conversion of time units as the BET model requires*) as a measuring time unit. This model represents *infinite-failure* models - permitting visual description of unlimited amount of failures [11]. In theory, both portray process in removing faults in software systems that comprises of *finite* number of faults.

An instance of using the LPET model is demonstrated in the Amazon system; there are over 304 million users, each user has the capacity to perform '*x*' amount of operations. The extent of operations per user (*x*) cannot be estimated as it varies but in theory it's not an *infinite* value – showing that even the most complex and utilized systems in essence have a finite number of possible failures. This model could then allow *software reliability engineering testing* (SRET) to focus on a specific calendar time, *i.e.* Load or Stress tests using data from Christmas periods as those are times of user engagements.

The difference in the two models is that the LPET model only uses actual data, where as the BET model may be used taking into account estimated failure data values.

III. DETERMINING RELIABILITY

Optimum reliability of software systems are beneficial in reducing expenditures during the life-cycle; defining a Test Plan with associated Test Cases help assess software reliability.

*A. Test Plan*

Ops a La Carte, a reliability engineering firm, states that a test plan must gather all SRE activities (*operational profiles, models, etc.*) to ensure testing needs are achieved (as specified in *step 3* of the SRE process and to also remove duplication of tests) [15]. Completing a well-thought SRET plan consists primarily of the operational profile (*refer to Appendix*) and the quantitative reliability objectives [5], *i.e. failure intensity objective – at what point should testing be stopped*. The use of operational profiles allows testing to be focused on critical functions related to the requirements of a system.

*1)* For the simplicity of understanding a test plan this report focuses on the pacemaker system aforementioned. Figures 8-8.2 illustrates the logical process of documenting a test plan, what it should define & contain.

*B. Test Case(s)*

As implied earlier a test plan is not thorough without one or more test cases, identified by IEEE Standard 610 as "a set of test inputs, execution conditions, and expected results developed for a particular objective, such as to exercise a particular program path or to verify compliance with a specific requirement" [16]. In effect, test cases must support discovery of information within software systems – each *run* (a particular instance of an operation) should document & focus on the following:

- Detected bugs
- Resolve bugs

For example, after altering code of pacemaker system *x* and measuring its performance, if over the span of 2 hours the performance rate has decreased there should be documentation tracing the bug to the newly inserted code. Additionally when code is altered to increase performance level over the same time unit, this should be documented.

SRE is an approach that takes "a global view of the product involved" [3], Failure Modes and Effects Analysis (FEMA) is another efficient approach "that looks at particular failures, how they can be caused, and how to prevent them" [3]. It is not recommended to use this approach as a primary reliability measuring method because of its cost (*requires detailed analysis of each operation and its variations*) but should supplement the SRE process after failure intensity has been established to then focus on failure prevention methods [3].

There are numerous testing types available for measuring reliability; the following were selected for the purpose of this report [3]:

- Functional – Purposed to test each feature of the system in isolation; reasonable to first focus on the operations documented in the operational profile then test interaction of numerous functions

- Load – Tests the system by constantly stimulating an abundance of users until the user threshold is met (loads the server with *dummy* users)

- Performance – Tests the speed of a system

- Regression – Used to test modifications in systems when a system undergoes changes; aims to reveal failures of older functions of system

- Scenario – Testing used to test hypothetical situations, helps analyze how a program deals with the simulated situation

- Stress – "Testing conducted to evaluate a system or component at or beyond the limits of its specified requirements with the goal of causing the system to fail" [17]

Figures 9–11 outline a sample test case for *pacemaker system x* which may be used to perform reliability calculation. The specified test cases reflect operations with a high probability of occurring as specified in the operational profile (Figure 4); this allows SRET efforts to focus on key functions within the system from a users perspective.

*1)* The test plan and test cases for *pacemaker system x* were chosen carefully to examine the objectives, reliability, and the overall goal of the system - providing a system that handles its threshold of users at any given time under any condition. The different test cases are not explicit to individual operations within the *testing operational profile*, but rather represent different variation each operation may undergo. Each variation effectively has a different test case; revealing different dimensions of failures and/or faults within each run.

### IV. CONCLUSION

This report documented methods of quantitatively measuring the reliability of a software system through use of the SRE process. According to Musa [5], software deployment and operational costs are believed to be the most affected by unreliable software. Systematic SRET supplements these aspects of software development from both a marketing and development standpoint. As implied in *Section III*, hardware testing and feature testing processes are insufficient when measuring reliability of a system due to its unprecedented techniques within industry.

In relation to the Agile Testing Quadrants (Figure 7), Q3 & Q4 represent methods of reliability testing as they focus on determining the robustness of a system; whereas testing within Q1 & Q2 represent primarily focus on featured testing. In which testing within Q2 may be initiated to form a foundation of specification for the system. Then performing testing such as *Load Testing* within Q4 to assess the robustness (reliability) of the system.

Through the usage of the SRE process, *development of an operational profile, segmenting operations,* and the like, testers are likely to establish reliability growth if followed appropriately. Nevertheless it is safe to say organizations and stakeholders are somewhat against the ideology of promoting individual testing focused on reliability of some operations, as the costs may seem *to* outweigh the benefits. In the case of *pacemaker system x*, the cost of meeting high standards of reliability is worth the cost as this system is detrimental to ones life. Stakeholders with the expertise of testers should determine this decision alongside an appropriate selection of a prediction model.

There are increasing developments of models in industry to better measure reliability. However, as the complexity of software systems increase – so will the definition of measuring reliability within these systems. Thus suggesting the prediction of software failures (reliability) to be promising yet an uphill task as technology improves.

# Appendix

| **Operation Initiators for** *pacemaker system x* |
|---|
| Doctor |
| Patient |
| System Administrator (Maintenance) |
| Communications Network |

Figure 2 – Operations initiators

| **Operations list for pacemaker system x** ||
|---|---|
| *Operation Initiator* | *Operation* |
| Doctor | -Add notification<br>-View statistics for a specified time frame<br>-Enter rhythm rate |
| Patient | -Add notification<br>-View statistics for a specified time frame |
| System Administrator | -Export data to warehouse |
| Communications Network | -View status of connectivity in specified location |

Figure 3 – Operations list

| *Pacemaker system's occurrence rates for pacemaker system x* |||
|---|---|---|
| *Operation* | *Occurrence rate (operations/hr)* | *Occurrence Probability* |
| View status of connectivity in specified location (within predefined range) | 6000 | 0.86330935251799 |
| Export data to warehouse | 600 | 0.0863309352518 |
| Enter rhythm rate | 100 | 0.01438848920863 |
| Add notification (Doctor) | 100 | 0.01438848920863 |
| View statistics for a specified time frame | 150 | 0.02158273381295 |
| Total | 6950 | 1 |

Figure 4 – Hypothetical occurrence measures for *pacemaker system x* (sample of 100)

- Unplanned Events
    - Crashes
    - Hangs
    - Functionally incorrect response
    - Untimely response – too fast or slow
- Planned Events
    - Updates requiring restart
    - Configuration changes requiring a restart
- Configuration failures
    - Application/System incompatibility error
    - Installation/setup failures

Figure 5 – Failure Classification

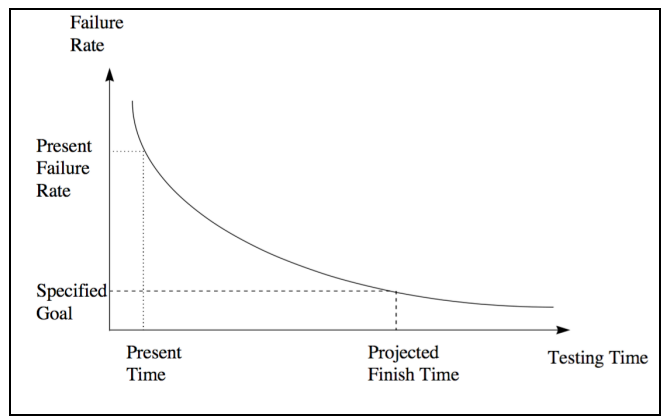

Figure 6 – Generic Software Reliability Model

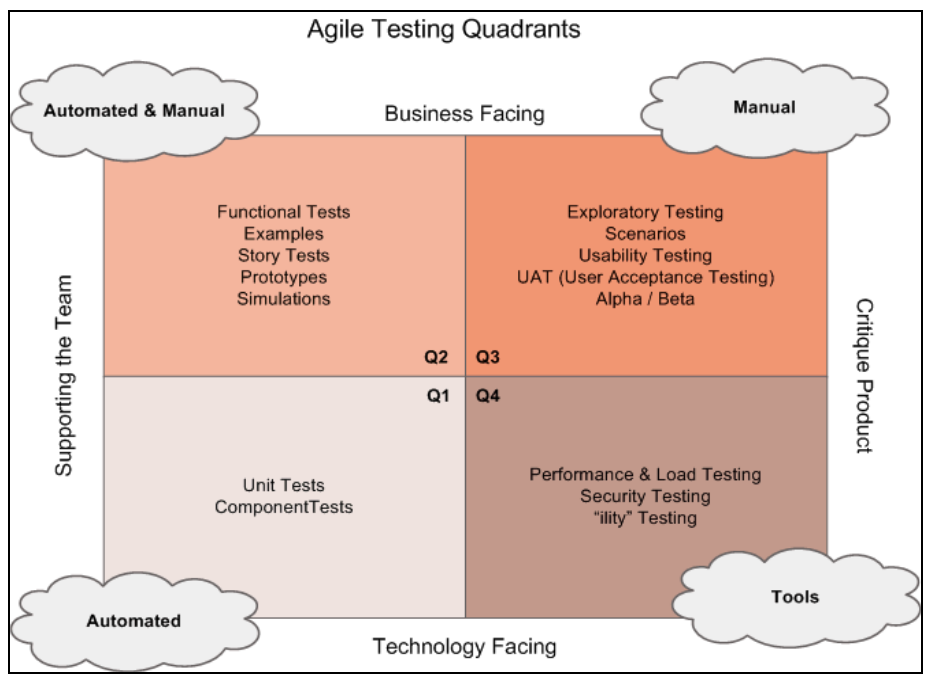

Figure 7 – Agile Testing Quadrants

The first objective one should consider when conducting a test plan is to determine which testing styles are appropriate for the system at hand. For the sake of *pacemaker system x*, the tests to be considered in measuring reliability as specified in Section III were due to the following reasons:

- Scenario Testing: System must be fully reliable regardless of environmental inputs such as being in basements with restricted connectivity, testing in hypothetical but unlikely environments will allow a better measurement of judging the system's reliability (*i.e. System being tested to see if pacemaker can update data to warehouse whilst inside a pool*)
- Load Testing: As there are several patients whom are continually updating their notification settings, doctors whom change the different heart rate settings, etc. rigorous testing should be undertaken to ensure that *pacemaker system x* is able to perform under mass amounts of functional requests
- Performance Testing: Similar reasoning as the *Load Testing*, though operations may be fully functional testing must be initiated to ensure speed of processing is optimized or discover areas where speed isn't optimized
- Functional Testing: Identifies individual components of the system (*reporting mechanisms, entering rhythm rate, etc.*) to ensure they are fully working at all times, then combines different functions (operations) together to assess their reliability once again
- Stress Testing: The *pacemaker system x* is one that may be utilized by millions in the future, this test allows measurement of robustness to determine how many patients/doctors will be able to fully use all features of the system at its present state
- Regression Testing: Though modification of code may not be current planned, it is advisable to consider regression testing to prepare validation of any modification that may occur due to bugs discovered in other forms of testing

After selecting the appropriate test types, gather the system's operations from the operational profile (Figure 4) to be tested and document the objectives each test should aim for. Figure 8 provides a high level view of the test objectives. The following must be accounted for:

- Reference: Numeric or alphabetical reference to each test, allows easy traceability
- Operation: Refers to those documented in the operational profile
- Test Objective: Objective that the test is to demonstrate
- Evaluation Criteria: Conditions to be evaluated to validate a successful test

| Test Reference | Operation | Test Objective | Evaluation Criteria |
|---|---|---|---|
| 1 | View statistics for a specified time frame | Aims to reveal if all pacemakers report statistical information when requested within time-constraint | All displayed statistics are accurate |
| 2 | Enter rhythm rate | Aims to reveal if user (doctor) can enter heart rhythm rate for a patient's pacemaker at any given time | Doctor is able to enter rhythm rate at any time in any environment |
| 3 | View status of connectivity in specified location | Aims to reveal if the pacemaker will be able to connect to the communications network at any given time, regardless of outside disturbance. | Device is connective to central system (data center) at all times |
| 4 | Add notification | Reveals if doctors and patients can add notifications (*message when heart rate is decreasing, etc.*) | Notification can be added by appropriate persons |
| 5 | Export data to warehouse | Reveals if System Administrators can export data to the warehouse at any given time | Determine if 100% of data is transferred to the warehouse |

Figure 8 – Test Plan Test Objectives Documentation

After documenting test objectives and evaluation criterion for each operation, a systematic strategy must entail of selecting test type according to the system's needs. Figure 8.1 provides an example of testing types and the objectives to be covered. For the simplicity of this report there will only be three examples.

| Test Type | Objectives |
|---|---|
| **Scenario Testing** | - Validate <u>Test Reference 3</u> under *unusual* operational environments |
| **Load Testing** | - Validate <u>Test Reference 5</u> by invoking multiple variations of this operation |
| **Performance Testing** | - Validate <u>Test Reference 1</u> by invoking a larger request than the system is use to handling |

Figure 8.1 – Test Type & Objectives

Following the documentation of tests to be carried out, test cases should be produced to provide detailed parameters within testing methods. Figures 9-11 illustrates sample cases taking into account the Test Types documented in Figure 8.1.

Figure 8.2 shows the tool(s) needed for testing per operation within the test case; tools should be documented as early as possible to ensure they are available during the SRET process. For simplicity of this report only one test case is considered.

| Test Case | Tool |
|---|---|
| Test Reference 5 | Load Runner |

Figure 8.2 – Testing Tools

| Test Case ID | **Test Reference 3** |
|---|---|
| **Description** | This test case aims to reveal if the pacemaker will be able to connect to the communications network at any given time, regardless of outside disturbance. |
| **Input** | **Direct**:<br>- Communications Network employee enters patients ID<br>- Communications Network employee enters ID for specific region<br>**Indirect**:<br>- Holiday season (*certain regions may have limited communications available due to shortage of staff*)<br>- Tornado storm (*causing network barriers*) |
| **Test Operations** | View status of connectivity in specified location (within predefined range) |
| **Failure Condition** | There are less than 100% of the pacemakers connected at anytime within the hour |
| **Expected Results** | All pacemakers to be connected to the system throughout the hour regardless of location. |
| **Actual Results** | Within the hour of testing, there were 4 pacemakers that did not maintain the function of being connected at all times. |
| **Time Started** | January 1, 2016 – 00:35 |
| **Time Finished** | January 1, 2016 – 01:35 |
| **Outcome of** | Failed - The test revealed that the some patients |
| test | pacemakers were not connected to the system at all times; plausible to say due to weather constraints. |

Figure 9 – Test Case Documentation

| | |
|---|---|
| **Test Case ID** | **Test Reference 5** |
| **Description** | This test should facilitate and measure if System Administrators can export data to the warehouse. |
| **Input** | **Direct**:<br>- System Administration inserts data warehouse location<br>- Pacemaker patient's ID<br>- Time frame of information to be extracted |
| **Test Operations** | Export data to warehouse |
| **Failure Condition** | System Administrators are not able to export data to warehouse. |
| **Expected Results** | System Administrators should be able to export pacemaker data to a specified warehouse. |
| **Actual Results** | System Administrators were able to export all pacemaker results to the data warehouse |
| **Time Started** | January 15$^{th}$, 2016 – 13:43 |
| **Time Finished** | January 15$^{th}$, 2016 – 14:43 |
| **Outcome of test** | Pass – All system administrators within the testing period were able to export all data into the specified data warehouse. |

Figure 10 – Test Case Documentation

| | |
|---|---|
| **Test Case ID** | **Test Reference 1** |
| **Description** | This test case should help measure and determine if a doctor can view the heart rate of a patient using the pacemaker. |
| **Input** | **Direct**:<br>- Doctor navigates to view statistics on user interface<br>- Doctor enters patient ID<br>**Indirect**:<br>- Patient with pacemaker is a pool/sauna |
| **Test Operations** | View statistics for a specified time frame |
| **Failure Condition** | Doctor is not able to view statistics of a patient within a specified time-frame. |
| **Expected Results** | Doctor should be able to view the statistics (*heart rate, temperature, etc.*) of patient with pacemaker. |
| **Actual Results** | Doctor was able to view statistics of patient accurately. |
| **Time Started** | February 14, 2016 – 17:45 |
| **Time Finished** | February 14, 2016 – 18:45 |
| **Outcome of test** | Pass – Doctors were able to view patients heart rate and other statistical information at any given time with the hour test |

Figure 11 – Test Case Documentation